\documentclass[fleqn,usenatbib]{mnras}

\usepackage{newtxtext,newtxmath}

\usepackage[T1]{fontenc}

\DeclareRobustCommand{\VAN}[3]{#2}
\let\VANthebibliography\thebibliography
\def\thebibliography{\DeclareRobustCommand{\VAN}[3]{##3}\VANthebibliography}

\usepackage{graphicx}	
\usepackage{amsmath}	
\usepackage{lineno}

\title[Gamma-ray attenuation in the {\it Fermi} and CTA era]{A new derivation of the Hubble constant from $\gamma$-ray attenuation using improved optical depths for the {\it Fermi} and CTA era}

\author[A. Dom\'inguez et al.]{
A. Dom\'inguez,$^{1}$\thanks{E-mail: alberto.d@ucm.es}
P. {\O}stergaard Kirkeberg,$^{2}$\thanks{E-mail: dtm518@alumni.ku.dk}
R. Wojtak,$^{2}$\thanks{E-mail: radek.wojtak@nbi.ku.dk}
A. Saldana-Lopez,$^{3}$\thanks{E-mail: alberto.saldanalopez@unige.ch}
A. Desai,$^{4}$\newauthor
J.~R. Primack,$^{5}$
J. Finke,$^{6}$
M. Ajello,$^{7}$
P.~G. P\'erez-Gonz\'alez,$^{8}$
V.~S. Paliya,$^{9}$
and D. Hartmann$^{7}$
\\
$^{1}$IPARCOS and Department of EMFTEL, Universidad Complutense de Madrid, E-28040 Madrid, Spain\\
$^{2}$DARK, Niels Bohr Institute, University of Copenhagen, Jagtvej 128, 2200 Copenhagen, Denmark\\
$^{3}$Department of Astronomy, University of Geneva, 51 Chemin Pegasi, 1290 Versoix, Switzerland\\
$^{4}$Dept. of Physics and Wisconsin IceCube Particle Astrophysics Center, University of Wisconsin–Madison, Madison, WI 53706, USA\\
$^{5}$Department of Physics, University of California, Santa Cruz, CA 95064, USA\\
$^{6}$U.S. Naval Research Laboratory, Code 7653, 4555 Overlook Avenue SW, Washington, DC 20375-5352, USA\\
$^{7}$Department of Physics and Astronomy, Clemson University, Kinard Lab of Physics, Clemson, SC 29634-0978, USA\\
$^{8}$Centro de Astrobiolog\'ia (CAB, CSIC-INTA), Carretera de Ajalvir km4, E-28850 Torrej\'on de Ardoz, Madrid, Spain\\
$^{9}$Inter-University Centre for Astronomy and Astrophysics (IUCAA), SPPU Campus, 411007, Pune, India
}

\date{Accepted XXX. Received YYY; in original form ZZZ}

\pubyear{2022}

\begin{document}
\label{firstpage}
\pagerange{\pageref{firstpage}--\pageref{lastpage}}
\maketitle

\begin{abstract}
We present $\gamma$-ray optical-depth calculations from a recently published extragalactic background light (EBL) model built from multiwavelength galaxy data from the {\it Hubble Space Telescope Cosmic Assembly Near-Infrared Deep Extragalactic Legacy Survey} (HST/CANDELS). CANDELS gathers one of the deepest and most complete observations of stellar and dust emissions in galaxies. This model resulted in a robust derivation of the evolving EBL spectral energy distribution up to $z\sim 6$, including the far-infrared peak. Therefore, the optical depths derived from this model will be useful for determining the attenuation of $\gamma$-ray photons coming from high-redshift sources, such as those detected by the Large Area Telescope on board the {\it Fermi Gamma-ray Space Telescope}, and for multi-TeV photons that will be detected from nearby sources by the future Cherenkov Telescope Array. From these newly calculated optical depths, we derive the cosmic $\gamma$-ray horizon and also measure the expansion rate and matter content of the Universe including an assessment of the impact of the EBL uncertainties. We find $H_{0}=62.4$~$^{+4.1}_{-3.9}$~km~s$^{-1}$~Mpc$^{-1}$ when fixing $\Omega_{m}=0.32$, and $H_{0}=65.1$~$^{+6.0}_{-4.9}$~km~s$^{-1}$~Mpc$^{-1}$ and $\Omega_{m}=0.19\pm 0.08$, when exploring these two parameters simultaneously.
\end{abstract}




\defcitealias{saldana-lopez21}{S21}

\section{Introduction}
Star-formation processes produce radiation that is redshifted and accumulated in intergalactic space over cosmic history. These photons at ultraviolet (UV), optical, and infrared (IR) wavelengths are known as the extragalactic background light (EBL), and its understanding is essential for building a complete picture of galaxy formation \citep[e.g.~][]{hauser98,driver08,dwek13,driver21}.

Gamma rays with energies above approximately 10~GeV traveling over cosmological lengths are attenuated through pair production by EBL photons \citep[e.g.~][]{nikishov62,gould66}. This interaction is dependent on the distance to the source and the energy of the $\gamma$-ray photon, leaving a signature in the $\gamma$-ray spectrum of extragalactic sources. This effect has been systematically measured with different instruments and techniques in blazars \citep[e.g.~][]{ebl12,hess_ebl13,dominguez13a,biteau15,hess_ebl17,ebl18,magic_ebl19,veritas_ebl19,desai19} and also $\gamma$-ray bursts \citep{desai17}.

An accurate and precise characterization of the EBL is necessary for the correct estimate of the intrinsic properties of blazars \citep[e.g.~][]{dominguez15,vandenberg19,nievasrosillo22}, the understanding of $\gamma$-ray propagation physics over space \citep[e.g.~][]{aharonian94,coppi97,deangelis07,sanchez-conde09,dominguez11b,broderick12,buehler20,franceschini21,biteau22}, the derivation of cosmological parameters \citep[][]{dominguez13b,biteau15,dominguez19} and also detectability predictions at TeV energies extrapolated from GeV observations \citep[e.g.~][]{hassan17,paiano21}.

However, most of current EBL models are constructed based on rather limited galaxy data \citep[e.g.~][]{finke10,dominguez11a,gilmore12,helgason12,stecker16,franceschini17,andrews17} resulting in large uncertainties, specially at high-redshift and in the mid-to-far IR. In order to reduce such uncertainties, we used a sample of more than 150,000 galaxies with multiwavelength observation from $z=0$ to $z=6$ to produce a purely observationally-based EBL model \citep[][hereafter \citetalias{saldana-lopez21}]{saldana-lopez21}. This model uses the unprecedented high-redshift and IR data from the {\it Cosmic Assembly Near-Infrared Deep Extragalactic Legacy Survey} (CANDELS) collaboration, thus reducing uncertainties and relying in less extrapolations than other previous empirical models. The CANDELS survey gathers the deepest datasets in the UV (from GALEX and HST/WFC3/UVIS), optical (from HST/ACS and ground based telescopes such as GTC, Subaru, or VLT), near-infrared (from HST/WFC3 and ground based telescopes such Subaru or UKIRT), mid-infrared (from Spitzer/IRAC), and far-infrared (from Spitzer/MIPS and Herschel PACS and SPIRE).

Combining measurements of the $\gamma$-ray attenuation and observationally determined EBL model enables estimating cosmological distance as a function of redshift. This method provides absolute calibration of cosmological distances and thus it is a self-contained technique to determine the Hubble constant \citep{dominguez19}. Measuring the Hubble constant based on diverse and independent methods of distance calibration will play an important role in pinning down the cause of the so-called Hubble constant tension \citep{Riess2022}. This problem is a discrepancy between the Hubble constant derived from the Planck observations of the cosmic microwave background (CMB) radiation \citep{planck19} and its counterpart obtained in the Supernovae and $H_{0}$ for the dark energy equation of state (SH0ES) program using observations of Cepheids and type Ia supernovae with direct local geometric distance anchors \citep{Riess2022}. The Hubble constant tension may signify a cosmological anomaly, which would require revision of the standard cosmological model, most likely in the early universe before recombination \citep{Schoenberg2021}. The leading cosmological solution to the Hubble constant tension involves early dark energy \citep{2023Poulin} and it has testable implications for structure formation at high redshifts \citep{klypin21}. On the other hand, the possibility of hidden systematic effects is not completely ruled out. For example, type Ia supernovae have been recently shown to be the most likely source of possible unaccounted systematic errors in the local determination of the Hubble constant \citep{Wojtak2022}.

In this article, we discuss the $\gamma$-ray optical depths, cosmic $\gamma$-ray horizon (CGRH), and the cosmological parameters from $\gamma$-ray attenuation extracted from \citetalias{saldana-lopez21}. This paper is organized as follows. In Section~\ref{sec:ebl}, we summarize the \citetalias{saldana-lopez21} model. Section~\ref{sec:attenuation} describes the theoretical aspects related with deriving the optical depths from an EBL model, including the cosmological dependence. Then, Section~\ref{sec:results} describes and discusses the results, followed by Section~\ref{sec:conclusions} that concludes this work.

\section{Extragalactic background light model}
\label{sec:ebl}
The \citetalias{saldana-lopez21} empirical EBL model relies on observationally derived galaxy spectral energy distributions (SEDs), which were used to constrain the evolution of the (galaxy) cosmic emissivity at 0.1-1000~$\mu m$ from $z = 0$ to 6, and thus the evolution of the EBL itself. This section summarizes the analysis and main results presented in \citetalias{saldana-lopez21}. 

\subsection{Galaxy data and spectral energy distributions}
In \citetalias{saldana-lopez21}, we used multiwavelength data from CANDELS \citep[][]{Grogin11, Koekmoer11}. CANDELS is a {\it Hubble Space Telescope} (HST) large program which made use of the Advanced Camera for Surveys (ACS) and the Wide Field Camera 3 (WFC3) to provide broad-band near-ultraviolet (NUV) to near-infrared (NIR) observations (0.4-1.6~$\mu m$). CANDELS observed in five of the so-called cosmological fields: GOODS-S \citep{Guo13}, UDS \citep{Galametz13}, COSMOS \citep{Nayyeri17}, EGS \citep{Stefanon17} and GOODS-N \citep[][]{Barro19}. Thanks to the use of five different fields, potential cosmic variance effects are mitigated in the \citetalias{saldana-lopez21} EBL determination. The effective sky area per field is $\sim$200~arcmin$^2$, with around 35,000 galaxies detected in each field. CANDELS includes two observational modes with different $5\sigma$ $H$-band limiting magnitude (m$_H^{lim}$): CANDELS/Wide (m$_H^{lim} \sim$ 27 AB) and CANDELS/Deep (m$_H^{lim} \sim$ 27.7 AB). Data from CANDELS have been also used for the determination of cosmological quantities such as the evolution of the star formation rate and stellar mass density of the Universe (via luminosity and mass functions) or the relative importance of dust-enshrouded star formation as a function of redshift.

The HST observations were augmented by a multitude of follow-up observations in other world-wide ground- and space-based facilities. In particular, the {\it Spitzer Infrarred Telescope} and the Herschel observatory provided valuable mid- and far-IR photometry (3.6-500~$\mu m$) for the five CANDELS footprints. Additionally, GALEX FUV and NUV observations are available for GOODS-S and GOODS-N fields \citep{Morrissey07}, and also UVIS data. Such a huge community effort results in more than 25 broad- and intermediate-band observations for each object in the CANDELS survey, one of the most comprehensive imaging data sets ever. 

To access the CANDELS data and retrieve the photometry for every galaxy, we used the Rainbow Cosmological Surveys Database\footnote{\url{http://arcoirix.cab.inta-csic.es/Rainbow_navigator_public}} \citep{B11a,B11b}. Our primary sample selection required detection in the F160W band, and a signal-to-noise threshold $\geq 5\sigma$ in the same band. After removing potential stars in the field \citep[cf.][]{Guo13}, the working sample is composed of 154,447 out of a total of 186,435 sources. The redshift of every source is given by the photometric redshift reported by the CANDELS collaboration.

The UV-to-NIR SED of every galaxy was modelled by stellar population templates and in a homogeneous fashion for the whole working sample, following \citet{Barro19}. The stellar population templates were extracted from a semi-empirical library of 1,876 synthetic SEDs, built from GOODS–S/N IRAC selected sources \citep{PG05,PG08}. Assuming a Chabrier initial mass function \citep{Chabrier2003} and a Calzetti extinction law \citep{Calzetti01}, the stellar emission of the reference templates was characterized using the \textsc{Pegase2.0} models \citep{PEGASE1997}, including nebular continuum and emission lines' flux contribution. The models were constructed assuming a single stellar population, driven by an exponentially declining star-formation history. 

Given the low number of detections at MIR-to-FIR bands, we followed a novel approach to infer the IR contribution for the galaxies in our sample. In short, we split the catalogs into three categories. For type (1), galaxies where there is detection by both {\it Spitzer}-MIPS/24$\mu m$ and at least one Herschel band, we take the best-fit \citet[][]{ChEl01} template to reproduce the IR-SED. In case (2), sources where the reddest detection is MIPS/24$\mu m$, the global dust emission has to be extrapolated from a single point. We use the expression from \citet{Wuyts11} that derives the total IR luminosity --L(TIR )-- at 8-1000$\mu m$ given the observed MIPS/24$\mu m$ flux density. Then, we associate a \citet{ChEl01} scaled dust emission template to every inferred L(TIR). Finally, in case (3), L(TIR) is computed from the SFR-excess at 2800\AA\ i.e., the difference between the SED-modelled SFR in the NUV and the dust-corrected SFR at the same wavelength, which gives an indication of the total energy absorbed by the dust \citep{schaerer13}. The corrected SFR at 2800\AA\ was computed following the methods in \citet{Barro19}, based on measured UV $\beta$-slopes and assuming typical literature IRX-$\beta$ relationships \citep{Meurer99}. Once again, a scaled \citet{ChEl01} model, recalibrated as a function of mass, was linked to every inferred L(TIR).

For more details about the CANDELS survey, sample selection and the derived galaxy-SEDs, see the dedicated sections in \citetalias{saldana-lopez21} and references therein. 

\subsection{Empirical formalism of the model} \label{sec:eblfor}
From the SED of every galaxy, in \citetalias{saldana-lopez21} we recovered the EBL SED by first calculating the evolution with redshift of the so-called cosmic luminosity density, $j(\lambda, z)$, at 0.1-1000~$\mu m$. 

The CANDELS sample was first divided into 15 redshift bins ($\Delta z_i$) from $z = 0$ to $z = 6$, and the `observed' luminosity density ($j_{\rm obs}(\lambda, \Delta z_i)$) obtained by stacking together the rest-frame galaxy-SEDs contained within the same redshift interval, and dividing by the corresponding comoving volume. Then, a completeness correction was applied to the `observed' luminosity density by adding the contribution of galaxies below the mass completeness limit at every $z$-bin. This contribution is computed using the \citet{Ilbert13} ($0 \leq z < 4$) and \citet{Grazian15} ($4 \leq z < 6$) Stellar Mass Functions, which gives us the number of `missing' galaxies as a function of redshift (i.e., below the mass limit). The characteristic emission of such populations of low-mass (or faint) galaxies is given by the average SED of the galaxies immediately above the mass-completeness limit, weighted by the fraction of star-forming versus quiescent galaxies \citep[UVJ selection, see ][]{Williams09, Whitaker11}. In S21, we noted that the effect of our completeness corrections is only noticeable at the higher redshifts of the model ($z \geq 3$). 

As a result, \citetalias{saldana-lopez21} obtained values for the monochromatic luminosity density (at specific band-passes), which are fully compatible with other reported values in the literature, in particular, at the FUV (1500\AA), NUV (2800\AA), $B$- (4400\AA) and $K$- (2.2~$\mu m$) bands, up to the higher redshifts available in the previous publications. Moreover, the derived total-IR luminosity density (integrated over at 8-1000$\mu m$) and the Cosmic Star-Formation History (CSFH) from the \citetalias{saldana-lopez21} model are in agreement with literature inferences up to $z=2$. At higher $z$, literature values usually fall below our estimations due to our acccounting for IR non-detected galaxies (see previous section). However, our derived CSFH is in agreement within $1\sigma$ with the $\gamma$-ray attenuation results by \citet{ebl18} whose method is, in principle, sensitive to all the EBL photons in the cosmic budget, giving us confidence regarding the validity of our model. 

Finally, the EBL-SED was computed by integrating the evolution of the comoving luminosity density, $j(\lambda, z)$, as follows \citep[see e.g.,][]{MoVdBoschWhite2010}, 
\begin{equation}\label{eq:EBL}
    \lambda I_{\lambda}(\lambda, z_i) = \dfrac{c^2}{4\pi\lambda} \int_{z_i}^{z_{max}} j\left(\lambda(1+z_i)/(1+z'), z')\right) \left|\dfrac{dt}{dz'}\right| dz',
\end{equation}

\noindent
where $\lambda I_{\lambda}(\lambda, z_i)$ is given in nW~m$^{-2}$ sr$^{-1}$, $c$ is the speed of light in vacuum, and the $|dt/dz'|$ factor comes from the adopted cosmology and is given by
\begin{equation}
    \label{dtdz}
    \frac{dt}{dz'}=\frac{1}{c(1+z')H(z')}.
\end{equation}
This cosmology is initially described by the fiducial flat $\Lambda$CDM framework with a matter density parameter $\Omega_{\rm m}=0.3$ and $H_{0}=70$~km~s$^{-1}$~Mpc$^{-1}$, although later we will let the cosmological parameters vary. For this cosmology, the Hubble parameter $H(z)$ is given by
\begin{equation}\label{eq:hubble}
    H^{2}(z)=H_{0}^{2}[\Omega_{\rm m}(1+z)^{3}+1-\Omega_{\rm m}]
\end{equation}
and the comoving volume $V_{\rm com}(z)=(4/3)\pi D_{\rm com}^{3}(z)$, where $D_{\rm com}(z)$ is the comoving distance. Uncertainties on the EBL model came from the proper propagation of single galaxy-SED errors into the luminosity density SED, and subsequently into the EBL-SED using the previous equation.

At $z = 0$, \citetalias{saldana-lopez21} report an integrated intensity of $23.1\pm 4.0$~nW~m$^{-2}$~sr$^{-1}$ for the optical peak, which is consistent within 1$\sigma$ with previous estimations from empirical models such as \citep[e.g.,~][]{finke10, dominguez11a, gilmore12, helgason12}, galaxy number counts such as those by \citet{driver16}, and the $\gamma$-ray derivations by \citet{ebl18,desai19}. For the IR peak, \citetalias{saldana-lopez21} give $32.0\pm7.6$~nW~m$^{-2}$~sr$^{-1}$, compatible with other galaxy counts studies such as \citet{madau00, driver16, bethermin12}. At higher redshifts, the evolution of the optical peak generally agrees with previous models up to $z=1$ and after that, our EBL model generally follows the lower bound of the the latest results from $\gamma$-ray attenuation in \citet{ebl18}, up to $z = 3$.  

We stress that for this current paper, the \citetalias{saldana-lopez21} EBL uncertainties are recalculated. For constructing the upper and lower regions of the uncertainty band, we took the maximum and minimum values of the luminosity densities at each redshift bin. This procedure also assumes that the uncertainties in each redshift bin are correlated and translates in an overestimate of the uncertainties. Now, the EBL uncertainties are recomputed by building 500 MCMC realizations of the EBL spectral intensities, and their evolution, that are permitted by the luminosity density measurements, drawing random values within the uncertainties. These are assumed to be independent thus they are estimated in 15 non-overlapping redshift bins.

In summary, the \citetalias{saldana-lopez21} EBL empirical model stands as the only one to date that provides the cosmic EBL intensity at $z = 0 - 6$ purely based on observational data, also capable to reproduce (1) a diversity of FUV, NUV, optical, NIR, and TIR luminosity density in the literature, (2) the CSFH, which is in agreement with usual galaxy evolutionary hierarchical scenarios, (3) the lower bound of the direct detection data, (4) galaxy counts, and (5) the latest results from $\gamma$-ray attenuation studies. These results will allow us to derive the evolution of the $\gamma$-ray opacity of the Universe with unprecedented accuracy. 

\section{Theoretical background on $\gamma$-ray attenuation}
\label{sec:attenuation}

The optical depth $\tau(E,z)$ that is produced by pair production interactions between $\gamma$ rays and EBL photons is analytically given by

\begin{equation}
\label{attenu}
\tau(E,z)=\int_{0}^{z} \Big(\frac{dl}{dz'}\Big) dz' \int_{0}^{2}d\mu \frac{\mu}{2}\int_{\varepsilon_{th}}^{\infty} d\varepsilon{'}\ \sigma_{\gamma\gamma}(\beta{'})n(\varepsilon{'},z'),
\end{equation}

\noindent where $E$ and $z$ are respectively the observed energy and redshift of the $\gamma$-ray source, $\sigma_{\gamma\gamma}$ is the photo-photon pair production cross section and $n(\varepsilon{'},z')$ is the proper number density of EBL photons with rest-frame energy $\varepsilon{'}$ at redshift $z'$ given by
\begin{equation}
    \label{num_density}
    \varepsilon{'}n(\varepsilon{'},z')=\frac{4\pi}{c\varepsilon{'}}\lambda I_{\lambda}(hc/\varepsilon{'},z').
\end{equation}
The cross section $\sigma_{\gamma\gamma}$ depends on relative rest-frame energies of $\gamma$-ray photon ($E'$), EBL photon ($\varepsilon{'}$) and the rest mass energy of electron $m_{e}c^2$ via
\begin{equation}
    \beta{'}=\frac{\varepsilon_{th}}{\varepsilon{'}},
\end{equation}
where $\varepsilon_{th}$ is the energy threshold for photon-photon pair production given by
\begin{equation}
\label{threshold}
\varepsilon_{th}\equiv \frac{2m_{e}^2c^{4}}{E'\mu}
\end{equation}
and variable $\mu=(1-\cos\theta)$ relates the energy threshold to the angle of interaction $\theta$.

The optical depth $\tau(E,z)$ given by Equation~\ref{attenu} depends on the cosmological model in a twofold way. First, the integral along the line of sight involves a relation between the proper distance $l$ and redshift $z$ given by
\begin{equation}
    \label{dldz}
    \frac{dl}{dz}=\frac{cdt}{dz}.
\end{equation}
Second, the cosmological model enters also calculation of the EBL photon density via comoving luminosity density $j(\lambda,z)$ and integral over lookback time given by Equation~\ref{eq:EBL}. Rederiving the EBL model in different cosmological models requires the following steps. The comoving luminosity density resulting from stacking SED templates in redshift bis is rescaled according to the following equation:
\begin{equation}
    \label{j_scaling}
    j=j_{\rm fid}\frac{D_{\rm L}(z_{i})^{2}}{D_{\rm L,fid}(z_{i})^{2}}\frac{V_{\rm com,fid}(z_{i}+\Delta z_{i}/2)-V_{\rm com,fid}(z_{i}-\Delta z_{i}/2)}{V_{\rm com}(z_{i}+\Delta z_{i}/2)-V_{\rm com}(z_{i}-\Delta z_{i}/2)},
\end{equation}
where $D_{\rm L}$ is the luminosity distance, $V_{\rm com}$ is the comoving volume, $z_{i}$ and $\Delta z_{i}$ is the middle point and width of the $i$-th redshift bin. Distances $D_{\rm L}(z_{i})$ and volumes $V_{\rm com}(z_{i})$ are computed in a new cosmological model, while their counterparts marked with the subscript "fid" are for the fiducial model assumed in \citetalias{saldana-lopez21}. The cosmological model also changes the stellar mass completeness limit and the corresponding completeness correction described by \citetalias{saldana-lopez21}. Here, the lower limit of stellar masses scales with $D_{\rm L}(z_{i})^{2}$, while the normalisation of the stellar mass function used to extrapolate galaxy counts below the completeness limit scales with the inverse comoving volume as $1/(V_{\rm com}(z_{i}+\Delta z_{i}/2)-V_{\rm com}(z_{i}-\Delta z_{i}/2))$.

Deriving the EBL density from observations of galaxies involves three consecutive scalings with the Hubble constant: $\propto H_{0}^{-2}$ for converting observed galaxy fluxes to luminosities, $\propto H_{0}^{3}$ for computing the comoving luminosity density and $\propto H_{0}^{-1}$ for integrating the luminosity density over redshifts, as expressed in Equation~\ref{eq:EBL}. These three scalings cancel each other out so that the EBL density becomes nearly independent of the Hubble constant and the estimated optical depth for $\gamma$-ray attenuation proportional to $H_{0}^{-1}$ via the integral along the line of sight (\ref{attenu}). Therefore, the Hubble constant can be directly constrained from independent measurements of the $\gamma$-ray attenuation in spectra of objects in the Hubble flow by fitting normalisation of the predictions based on EBL model. Additional constraints from analogous measurements at high redshift are degenerated with the matter density parameter. This degeneracy results primarily from fitting a cosmological model to effectively a single distance measurement and it can be broken to a large extent by measuring the $\gamma$-ray attenuation in a possibly wide range of redshifts.

\begin{figure*}
	\includegraphics[width=2\columnwidth]{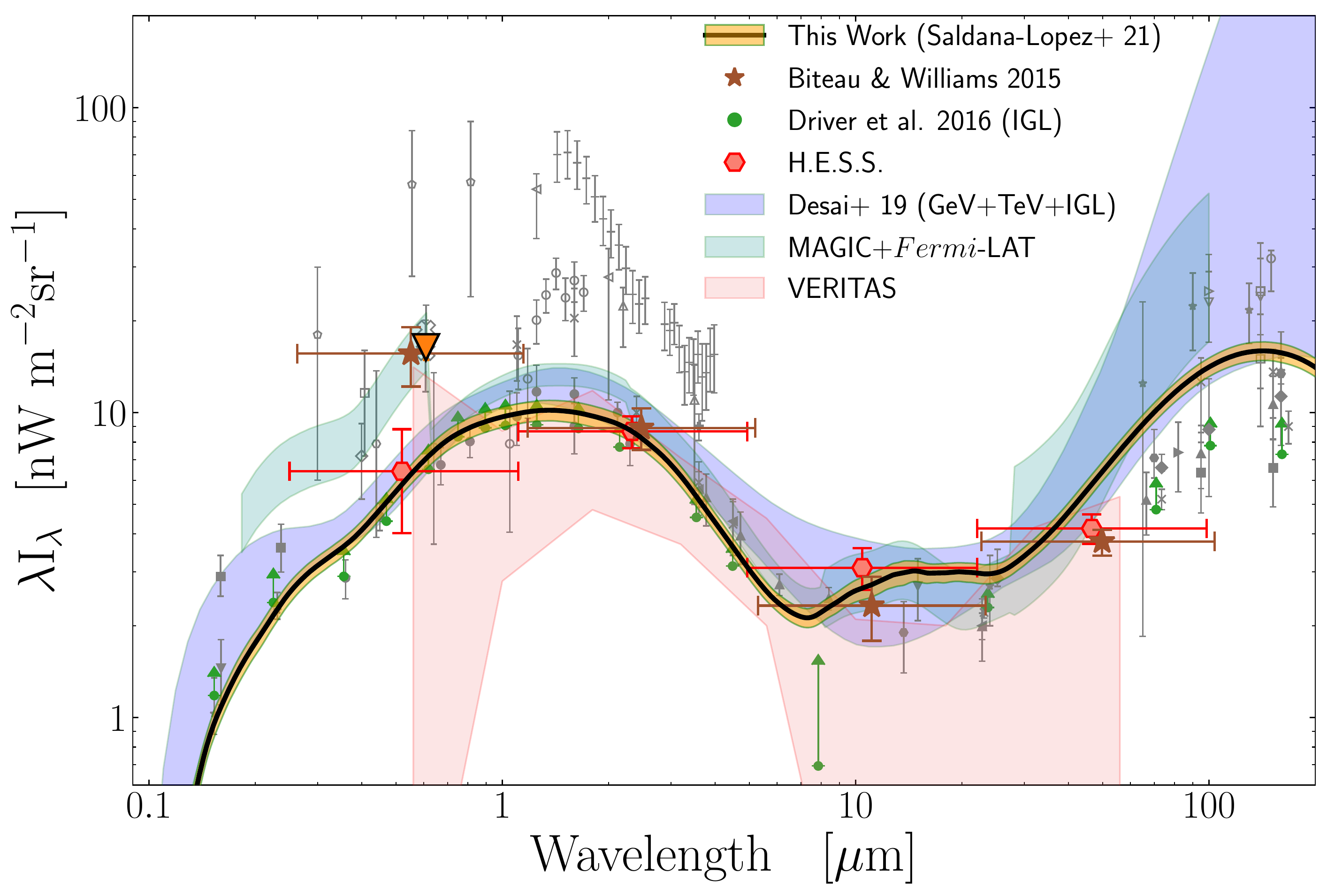}
    \caption{Spectral intensities of the extragalactic background light at $z=0$ from the EBL model by \citetalias{saldana-lopez21} (orange band, $1\sigma$ uncertainty), galaxy counts from \citet[][green symbols]{driver16}, and results based on $\gamma$-ray data from \citet[][brown symbols]{biteau15}, the H.E.S.S. collaboration in \citet[][red symbols]{hess_ebl17}, the MAGIC collaboration in \citet[][turquoise band]{magic_ebl19} showing here their results using the D11 model, the VERITAS collaboration using their 68\% containment result by \citet[][light red band]{veritas_ebl19} and \citet[][purple band]{desai19}. We refer to the reader to the main text for an explanation on how the systematic uncertainties affect these results. The direct detection data from \citet{lauer22} from {\it New Horizons} are shown at $\lambda=0.608$~$\mu$m. (orange triangle, uncertainties for this are barely visible). Note that the uncertainties that we are showing for the \citetalias{saldana-lopez21} model are smaller than the ones in that paper because we have recalculated them using a more appropriate methodology that is described in Section~\ref{sec:eblfor}.}
    \label{fig:ebl}
\end{figure*}

\section{Results and discussion}
\label{sec:results}
In this section, we compare the EBL estimates from \citetalias{saldana-lopez21} with $\gamma$-ray attenuation data, and we present and discuss the results from the optical depths calculations, cosmic $\gamma$-ray horizon, and the measurement of $H_{0}$ and $\Omega_{m}$.

\subsection{Extragalactic background light comparison with other $\gamma$-ray attenuation data } \label{sec:4.1}
The EBL spectral energy distribution in the local Universe from \citetalias{saldana-lopez21} is shown in Figure~\ref{fig:ebl} (note that the fiducial values for $H_{0}$ and $\Omega_{m}$ are used). This figure also shows the results derived from $\gamma$-ray attenuation data and galaxy counts. Note that the results determined from galaxy data tend to be on the lower bound of the calculations by \citet{desai19}. We note the intriguing discrepancy at about 4$\sigma$ from the absolute photometry measurement by the {\it New Horizons} spacecraft \citep{lauer22} and intensities derived from galaxies \citep[e.g.,~][]{driver16} at approximately $0.6\mu$m \citep[see also][]{symons23}. We stress that in Figure~\ref{fig:ebl} the $\gamma$-ray results by the MAGIC and VERITAS collaborations \citep[][respectively]{magic_ebl19,veritas_ebl19} are compatible with the \citet{lauer22} data. The results from MAGIC at these wavelengths become upper limits when including systematic uncertainties. The same happens to the measurements by \citet{biteau15}. Furthermore, we also note that the results from the H.E.S.S. collaboration \citep{hess_ebl17} when including systematic uncertainties are compatible with \citet{biteau15}. Therefore, in general, the $\gamma$-ray data from IACTs cannot rule out the high intensity EBL obtained by \citet{lauer22}.

\begin{figure*}
	\includegraphics[width=2\columnwidth]{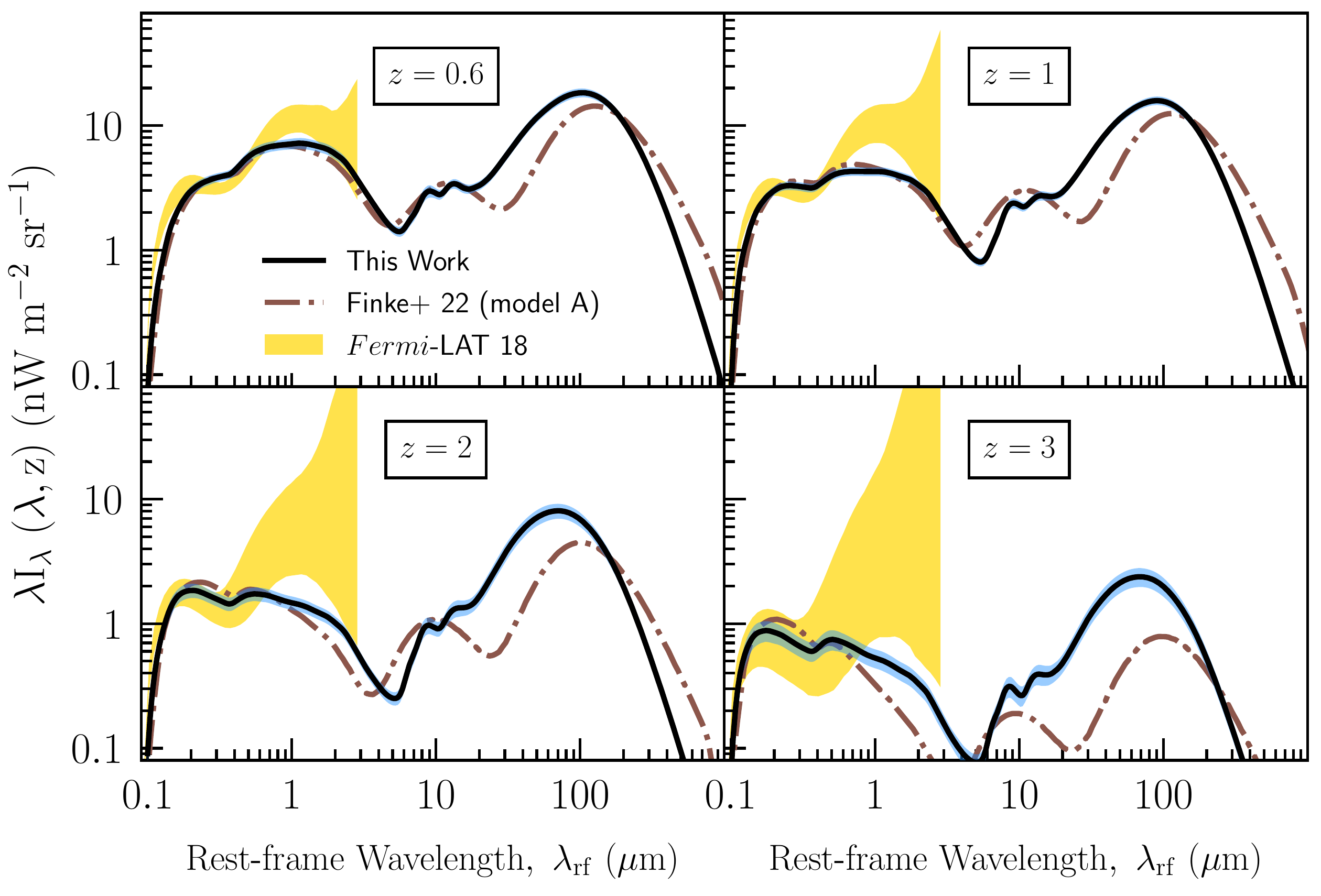}
    \caption{Spectral intensities of the extragalactic background light in four different redshits in co-moving frame from the EBL model by \citetalias{saldana-lopez21} (blue band, $1\sigma$ uncertainty), \citet{finke22} (brown line) and those from \citet[][yellow band]{ebl18}.}
    \label{fig:ebl_evol}
\end{figure*}

Figure~\ref{fig:ebl_evol} shows the evolution of the EBL from \citetalias{saldana-lopez21}, the new model by \citet{finke22} based on fitting a large compilation of galaxy data, and $\gamma$-ray attenuation data from \citet{ebl18} at four different redshifts. We see some discrepancies between the results derived from galaxies and $\gamma$-ray attenuation in the near infrared wavelength range at high redshifts $z\gtrsim 0.6$. It is not entirely clear what is the source of this tension. One possibility is that there exists an extra amount of the EBL, which is not accounted for in the current EBL models. The additional EBL component would dominate in the near infrared range and perhaps at longer wavelengths. On this direction, there are works claiming that there may be a significant amount of light from galaxy halos, known as intra-halo light or from low-surface brightness galaxies \citep[e.g.,~][]{zemcov2014,borlaff19,trujillo21,cheng21}. Another reason for the discrepancy may lie in the measurements of the $\gamma$-ray attenuation. In this case, overestimation of the EBL density derived from the $\gamma$-ray attenuation data may be caused by a hidden systematic effect biasing up the optical depth measurements at high energies and redshifts. This tension could be explained by extra effects in the $\gamma$-ray photon propagation such as the development of particle cascades \citep[e.g.~][]{coppi97} with or without the phenomenon of plasma instabilities \cite[e.g.~][]{broderick12}, although there is no other evidence of this effect so far \citep[e.g.,~][]{finke15,ackermann18}. There could be also more exotic solutions such as the existence of axion-like particles \citep[e.g.~][]{deangelis07,sanchez-conde09,buehler20} or dark matter decay \citep[e.g.~][]{kalashev19,bernal22,carenza23}.



\subsection{Optical depths}
\label{sec:opticaldepths}
\begin{figure*}
	\includegraphics[width=\columnwidth]{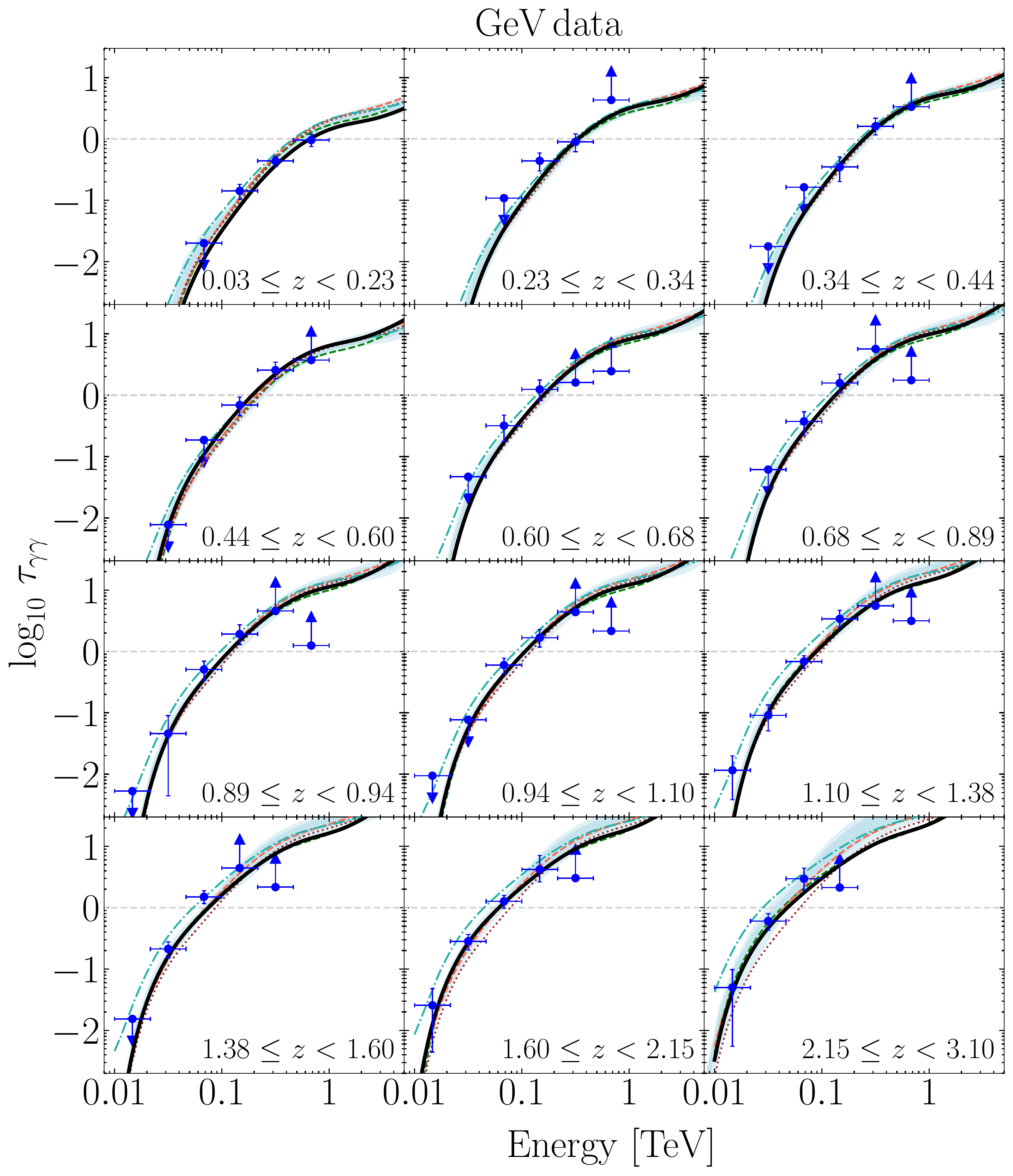}
	\includegraphics[width=5cm]{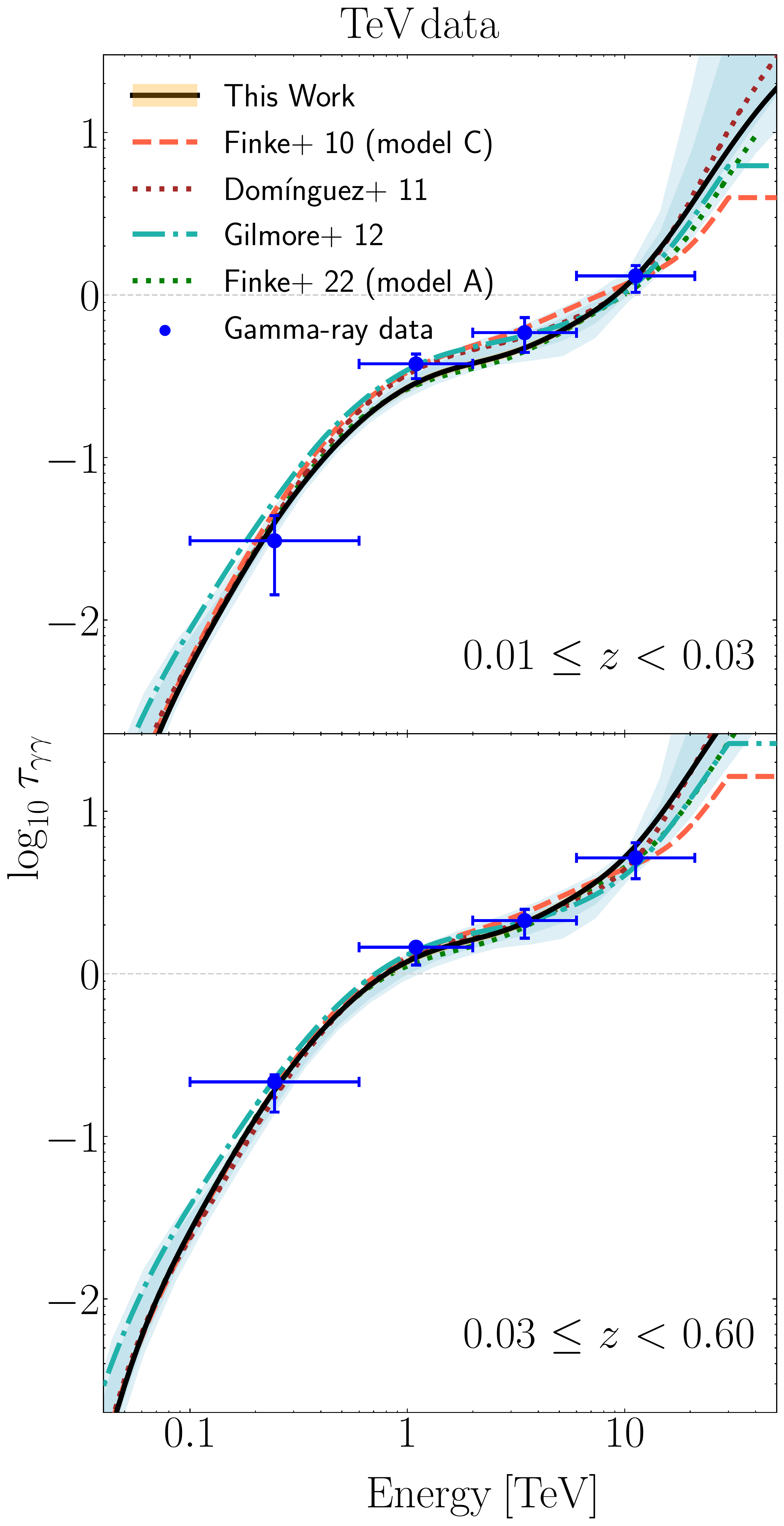}
    \caption{Optical depths as a function of energy in different redshift bins from EBL models \citep{finke10,dominguez11a,gilmore12,saldana-lopez21,finke22} and extracted from {\it Fermi}-LAT data \citep[{\it Left panel},][]{ebl18} and data from Imaging Atmospheric Cherenkov Telescopes \citep[{\it Right panel},][]{desai19}. The uncertainties from this work are small and barely visible. In both panels, 1$\sigma$ (dark cyan) and 2$\sigma$ (light cyan) uncertainties from \citet{ebl18} and \citet{desai19}, respectively, are also shown.}
    \label{fig:od}
\end{figure*}

By substituting $n(\varepsilon,z)$ in Equation~\ref{attenu} for that given by \citetalias{saldana-lopez21}, we can estimate optical depths\footnote{These files are available at \url{https://www.ucm.es/blazars/ebl}}. Figure~\ref{fig:od} shows these calculations in comparison with those from other EBL models and $\gamma$-ray attenuation data. We note that relative to other models, the agreement is good for the lower redshifts but our results tend to estimate lower optical depths at the higher redshifts, in particular for large energies of the order of 1~TeV. In comparison with the $\gamma$-ray attenuation data \citep{ebl18,desai19}, our results are in general in rather good agreement within uncertainties.

\subsection{Cosmic $\gamma$-ray horizon}
\label{sec:CGRH}
\begin{figure}
	\includegraphics[width=\columnwidth]{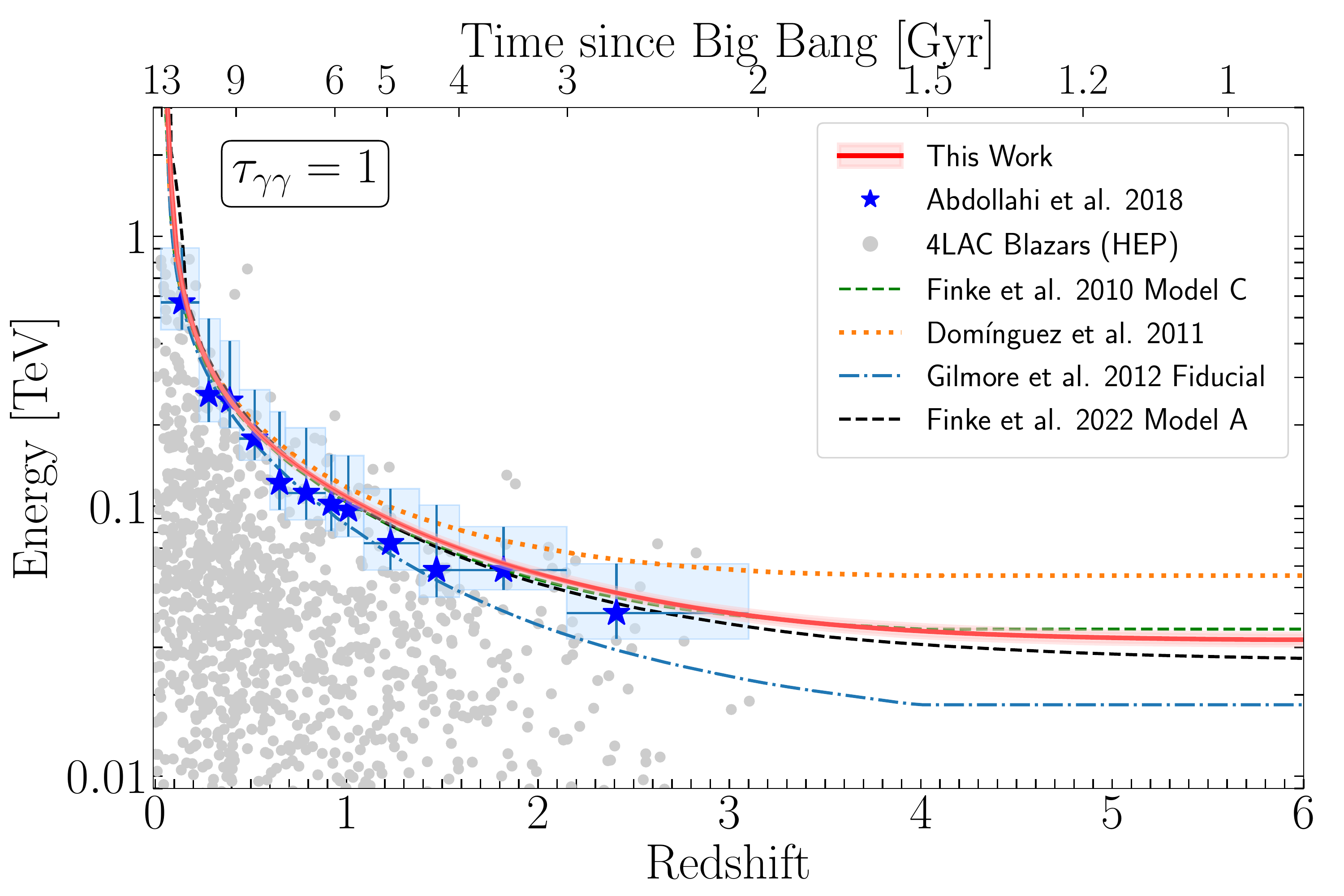}
    \caption{The CGRH from the EBL models by \citet[][green-dotted line]{finke10}, \citet[][orange-dotted line]{dominguez11a}, \citet[][brown-dotted line]{helgason12}, \citet[][cyan-dashed line]{gilmore12}, which have been extrapolated up to $z=6$, and this work (red band) from observations up to $z=6$, including uncertainties. We show the CGRH derived from $\gamma$-ray blazars \citep[][blue stars]{ebl18} and the highest energy photons detected for 4LAC blazars \citep[][grey circles]{4LAC}.}
    \label{fig:CGRH}
\end{figure}

An important cosmological observable is the energy as a function of the distance (or redshift) at which the optical depth is one. This property of the Universe is known as the CGRH and limits the Universe into a transparent and opaque region \citep[e.g.~][]{fazio70,dominguez13a}. As an example, looking at Figure~\ref{fig:CGRH}, it is less likely to observe 1~TeV photons from $z>0.1$ since those photons will be strongly attenuated. The same is expected for 100~GeV photons coming from $z>1$. However the Universe is transparent for photons with energies $E<10$~GeV. We stress that the estimate from the \citetalias{saldana-lopez21} model is in agreement with the independent measurement from \citet{ebl18}.

Note that there are photons on the more opaque region of Figure~\ref{fig:CGRH}. This is not surprising because the $\gamma$-ray attenuation is a probabilistic effect. Furthermore, in this figure, there are some effects that may bias its interpretation such as the exposure time, the blazars intrinsic spectrum, or their variability. For more context, consider that in the 4LAC-DR3 catalog \citep{4FGL-DR3,4LAC-DR3}, there are approximately 583 BL Lac sources with no redshift information, and about 120 of these have been observed with highest energy photons exceeding 100 GeV. As these BL Lac sources lack measured redshifts, they are not included in Figure~\ref{fig:CGRH}. We also checked the two sources with largest optical depths, these are, 4FGL J0507.9+6737 ($\tau=3.7$) and 4FGL J1224.1+2239 ($\tau=5.3$) at $z=0.42$ and $z=0.48$, respectively (redshifts given by the {\it Fermi}-LAT collaboration). From our search, we find that the redshift of 4FGL J0507.9+6737 is given as $z=0.314$ by \citet{mao11} and also photometrically by \citet{3hsp} as $z=0.34$, flagged as uncertain. For 4FGL J1224.1+2239, the redshift comes from a photometric estimate quoted by \citet{foschini22} as $z=0.479\pm 0.151$. In both cases, the redshift uncertainty brings down significantly $\tau$.

The possible derivation of new physics from the interpretation of Figure~\ref{fig:CGRH} is appealing, however caution should be taken since redshifts measurements for blazars, especially for BL Lac sources, can be challenging \citep[e.g.,~][]{rajagopal20,olmo-garcia22}, plus statistical fluctuations and other effects mentioned above can also play a role.

\subsection{The local Hubble constant and matter density of the Universe}

\label{sec:Hubble}

We constrain the Hubble constant $H_{0}$ \citep[see][]{salamon94} and the matter density parameter $\Omega_{\rm m}$ by fitting optical depths based on the EBL model from \citetalias{saldana-lopez21} to the $\gamma$-ray measurements \citep{ebl18,desai19}. A flat $\Lambda$CDM cosmological model is assumed (see Equation~\ref{eq:hubble}) and the dependence on cosmology both in the EBL model and the line-of-sight integral for the optical depth is taken into account. We use log-likelihood given by
\begin{equation}
    \ln L\propto -\sum_{i,j} \frac{[\tau_{\rm model}(z_{i},E_{j})-\tau_{i,j}]^{2}}{2\sigma_{i,j}^2},
\label{likelihood}
\end{equation}
where indices $i$ and $j$ run over redshift and energy bins, $\tau_{i,j}$ are the measured values of the optical depth and $\sigma_{i,j}$ takes the upper or lower error if respectively $\tau_{\rm model}(z_{i},E_{j})>\tau_{i,j}$ or $\tau_{\rm model}(z_{i},E_{j})<\tau_{i,j}$. For the upper and lower limits we use $\sigma_{i,j}\gg1$. Best fit cosmological parameters and the corresponding uncertainties are computed using the Markov Chain Monte Carlo method implemented in the Python-based \textit{emcee} code \citep{emcee}.

The likelihood function given by Equation~\ref{likelihood} is conditioned to a fixed EBL model and thus does account for errors in the EBL model. Incorporating uncertainties of the EBL model directly in the likelihood would require computations of rather non-trivial correlations between redshift and energy bins of the optical depth. Instead, we account for the EBL uncertainties in our cosmological analysis by marginalising it over EBL spectral intensities permitted by its measurement errors. The marginalisation is carried out by recomputing MCMC over a large number of Monte Carlo realisation (500) of the EBL model and finding best fit cosmological parameter using the sum of all chains. We calculate the EBL realisations by drawing samples from the luminosity density given its errors and computing the corresponding EBL densities. Uncertainties of the luminosity function take into account errors of multi-band photometry and extrapolations of the stellar mass function to its low-mass end. They are estimated in 15 non-overlapping redshift bins and thus assumed to be independent. 


\begin{table*}
    \centering
    \begin{tabular}{lcc}
    Data & $H_{0}$[km~s$^{-1}$~Mpc$^{-1}$] & $\Omega_{\rm m}$ \\
    \hline
    \hline
       &  & \\
    $\tau$ & $65.1^{+6.0}_{-4.9}$ &$0.19\pm{0.08}$ or $(<0.35\; (95\%))$ \\
      &  & \\
    $\tau(z<0.68)$     & $66.0^{+5.7}_{-5.1}$ & $<0.8\; (95\%)$ or  $(0.3^{+0.3}_{-0.2})$ \\
       &  & \\
    $\tau$+BAO(BBN) & $66.5^{+2.2}_{-2.1}$ & $0.28\pm{0.04}$\\
       &  & \\
    $\tau(z<0.68)$+BAO(BBN) & $69.8^{+3.0}_{-2.6}$ &  $0.34\pm{0.04}$ \\
     & & \\
    BAO(BBN) & $73.7\pm3.7$ & $0.396\pm0.054$\\
    \end{tabular}
    \caption{Constraints on the Hubble constant $H_{0}$ and the matter density 
    parameter $\Omega_{\rm m}$ from our joint analysis of independent $\gamma$-ray attenuation measurements. Beside our fiducial fit on the top, we include other fits restricted to $z<0.68$ $\gamma$-ray attenuation data ($\tau(z<0.68)$), also including BAO observations with the BBN prior as an external data set and only BAO observations with the BBN prior. Best fit parameters are provided as the posterior mean values with errors given by 16th and 84th percentile of the marginalised probability distributions. For $\Omega_{\rm m}$ inferred solely from the $\gamma$-ray data we provide 95 per cent upper limits. The errors include uncertainties in both the $\gamma$-ray observations and the EBL model.}
    \label{tab:cosmo-params}
\end{table*}

The optical depth data are taken from \citet{ebl18} and \citet[][see these references for details, in particular, Figure~2 of the latter one]{desai19}. 
In \citet{ebl18}, optical depths are estimated by measuring the $\gamma$-ray attenuation from a sample of 739 blazars plus one $\gamma$-ray burst, all detected by {\it Fermi}-LAT. These optical depths are given in twelve redshifts bins reaching $z\sim 3.10$. These redshift bins are chosen in such a way that the signal's strength is the same in each one of them. The optical depths are given in six logarithmically equally spaced energy bins from approximately 10~GeV up to 1000~GeV. The \citet{ebl18} results are especially relevant for constraining $\Omega_{m}$ because the larger dependence of the optical depth with $\Omega_{m}$ occurs at the larger redshifts. In \citet{desai19}, a sample of 38 blazars detected by Imaging Atmospheric Cherenkov Telescopes was used to measure the optical depths in two redshift bins up to $z\sim 0.6$. These optical depths are measured in four equally spaced logarithmic energy bins from 0.1~TeV up to approximately 20~TeV. These results from \citet{desai19} are especially important for measuring $H_{0}$ because the largest dependence of the optical depth with $H_{0}$ occurs at the higher energies and lower redshifts as shown by \citet{dominguez19}. We stress, as stated by \citet{ebl18} and \citet{desai19}, apart from the statistical uncertainty, an additional $\approx10\%$ systematic uncertainty exists. This systematic uncertainty accounts for shape differences of optical depth curves, intrinsic $\gamma$-ray spectral models, and energy biases. The impact of these uncertainties are taken into account and the combined statistical plus systematic uncertainties on the EBL optical-depth estimates are reported by these works.


\begin{figure}
	\includegraphics[width=\columnwidth]{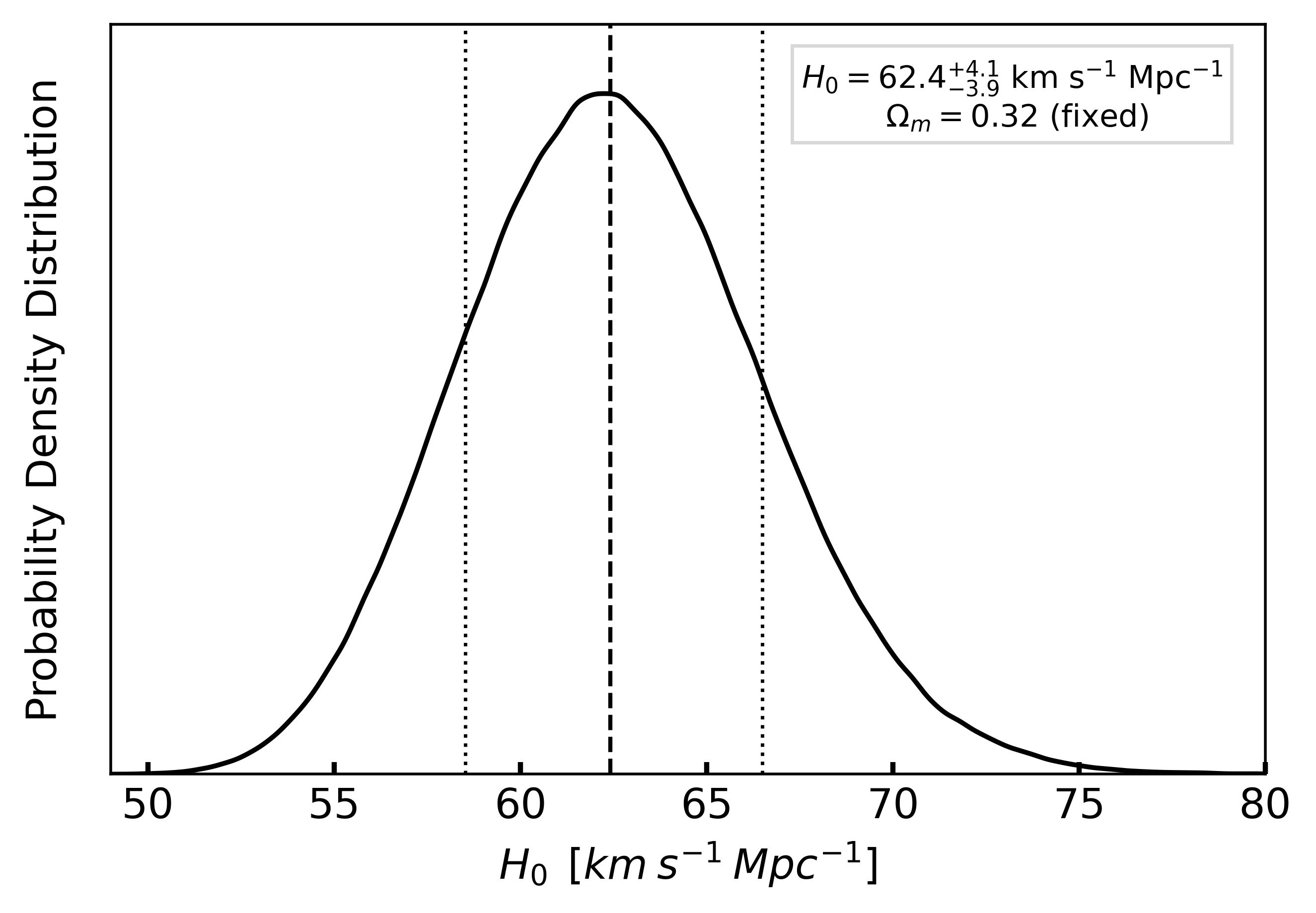}
    \caption{Constraints on the Hubble constant assuming $\Omega_{m}=0.32$. We obtain $H_{0}=62.4$~$^{+4.1}_{-3.9}$~km~s$^{-1}$~Mpc$^{-1}$.}
    \label{fig:H0}
\end{figure}

\begin{figure}
	\includegraphics[width=\columnwidth]{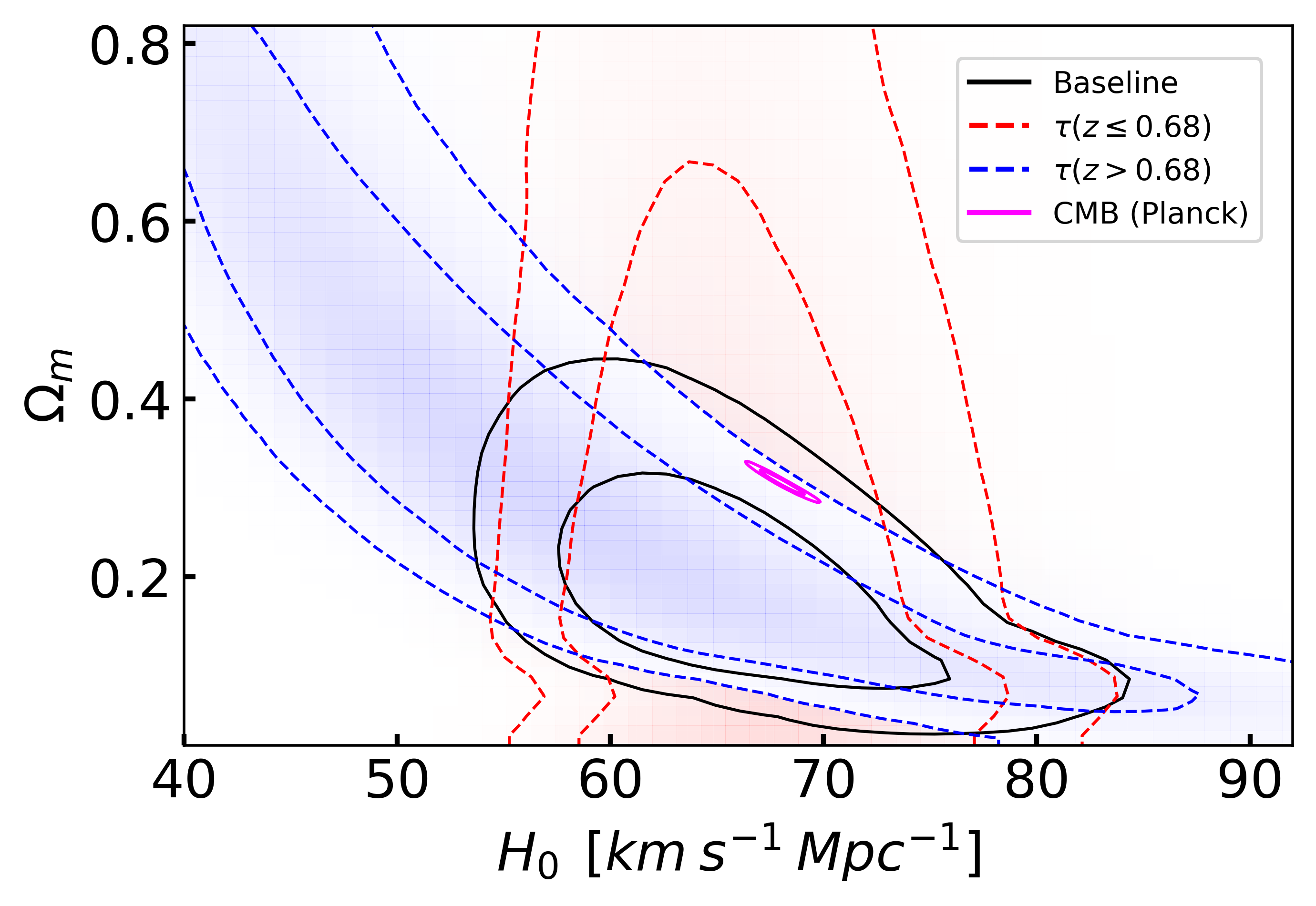}
    \caption{Constraints on the Hubble constant $H_{0}$ and the matter density parameter $\Omega_{\rm m}$ inferred from the measurements of $\gamma$-ray attenuation and the EBL model from \citetalias{saldana-lopez21}. Derived cosmological parameters are consistent with the Planck
    cosmological model when fits are limited to low-redshift ($z<0.68$) data, but only marginally consistent with the Planck model when we include high-redshift ($z>0.68$) data, for which low values of $\Omega_{\rm m}$ are required to match the optical depth based on the EBL model against $\gamma$-ray measurements. The contours show $1\sigma$ and $2\sigma$ confidence regions containing 68 and 95 per cent of the marginalised probability distributions.}
    \label{fig:cosmo-gamma}
\end{figure}

Our methodology leads to $H_{0}=62.4^{+4.1}_{-3.9}$~km~s$^{-1}$~Mpc$^{-1}$ when we fix $\Omega_{m}=0.32$, shown in Figure~\ref{fig:H0}, as found by \citet{aghanim18}. Figure~\ref{fig:cosmo-gamma} shows the main cosmological constraints inferred from the complete $\gamma$-ray data set, exploring simultaneously the $H_{0}$ and $\Omega_{m}$ parameter space. In this case,  the best fit Hubble constant is $H_{0}=65.1^{+6.0}_{-4.9}\,\rm{km}\,\rm{s}^{-1}\,\rm{Mpc}^{-1}$ and it is consistent with its value derived from the Planck observations of the CMB \citep{planck19} assuming the standard flat $\Lambda$CDM cosmological model. The data favour lower values of the matter density parameter than $\Omega_{\rm m}\approx0.3$ measured consistently from a wide range of observations \citep[e.g.,~][]{scolnic18}. However, this trend is not strong enough to be considered discrepant with other cosmological probes. In particular, $\Omega_{\rm m}$ from the Planck cosmological model lies well within $2\sigma$ credibility range of our constraints.

\begin{figure}
	\includegraphics[width=\columnwidth]{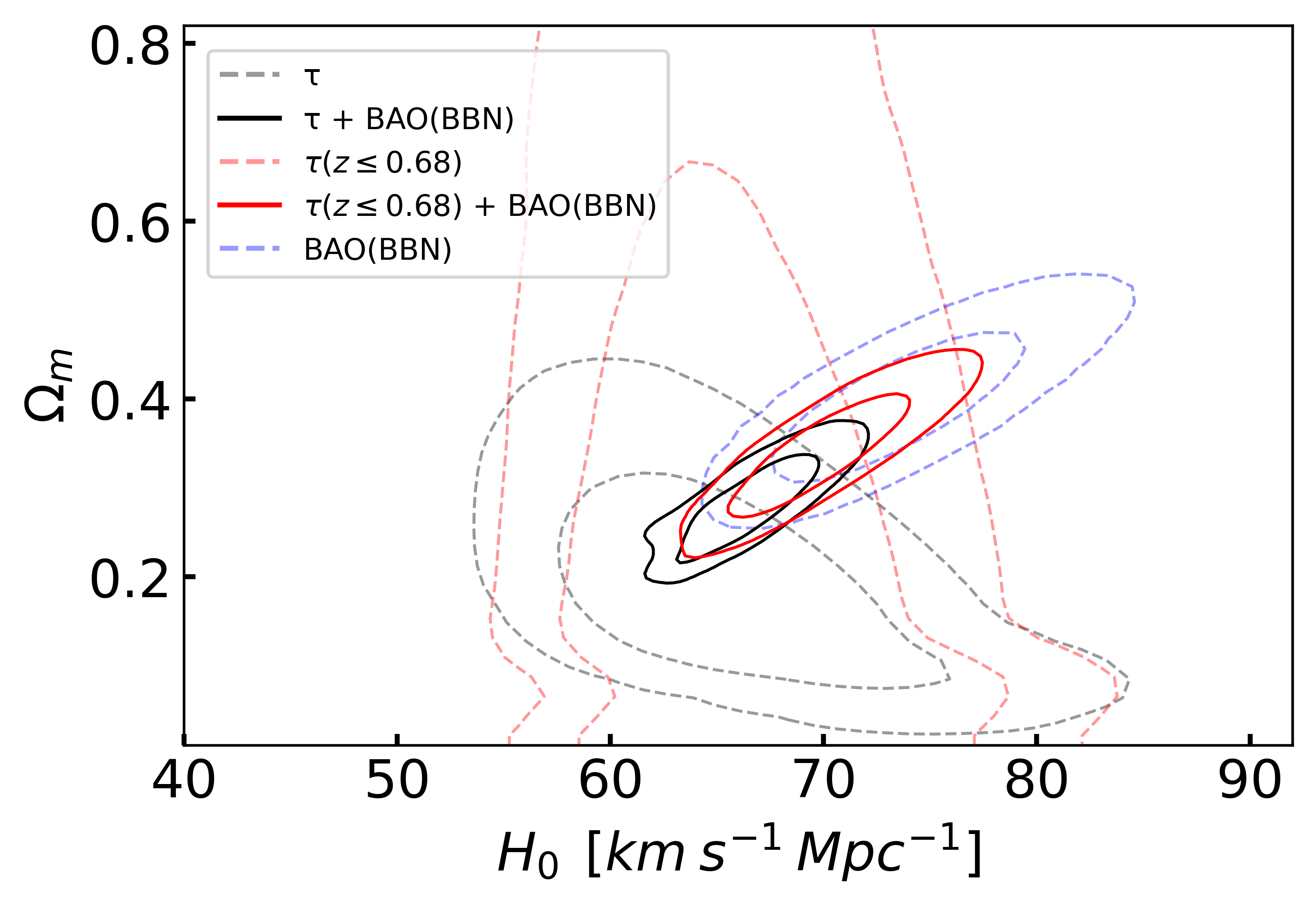}
    \caption{Constraints on the Hubble constant $H_{0}$ and the matter density parameter $\Omega_{\rm m}$ from a joint analysis of the $\gamma$-ray attenuation measurements and 
    Baryon Acoustic Oscillation (BAO) observations with the Big Bang Nucleosynthesis (BBN) prior. The contours show $1\sigma$ and $2\sigma$ confidence regions containing 
    68 and 95 per cent of the marginalised probability distributions.}
    \label{fig:cosmo-gamma-bao}
\end{figure}

\begin{figure}
	\includegraphics[width=\columnwidth]{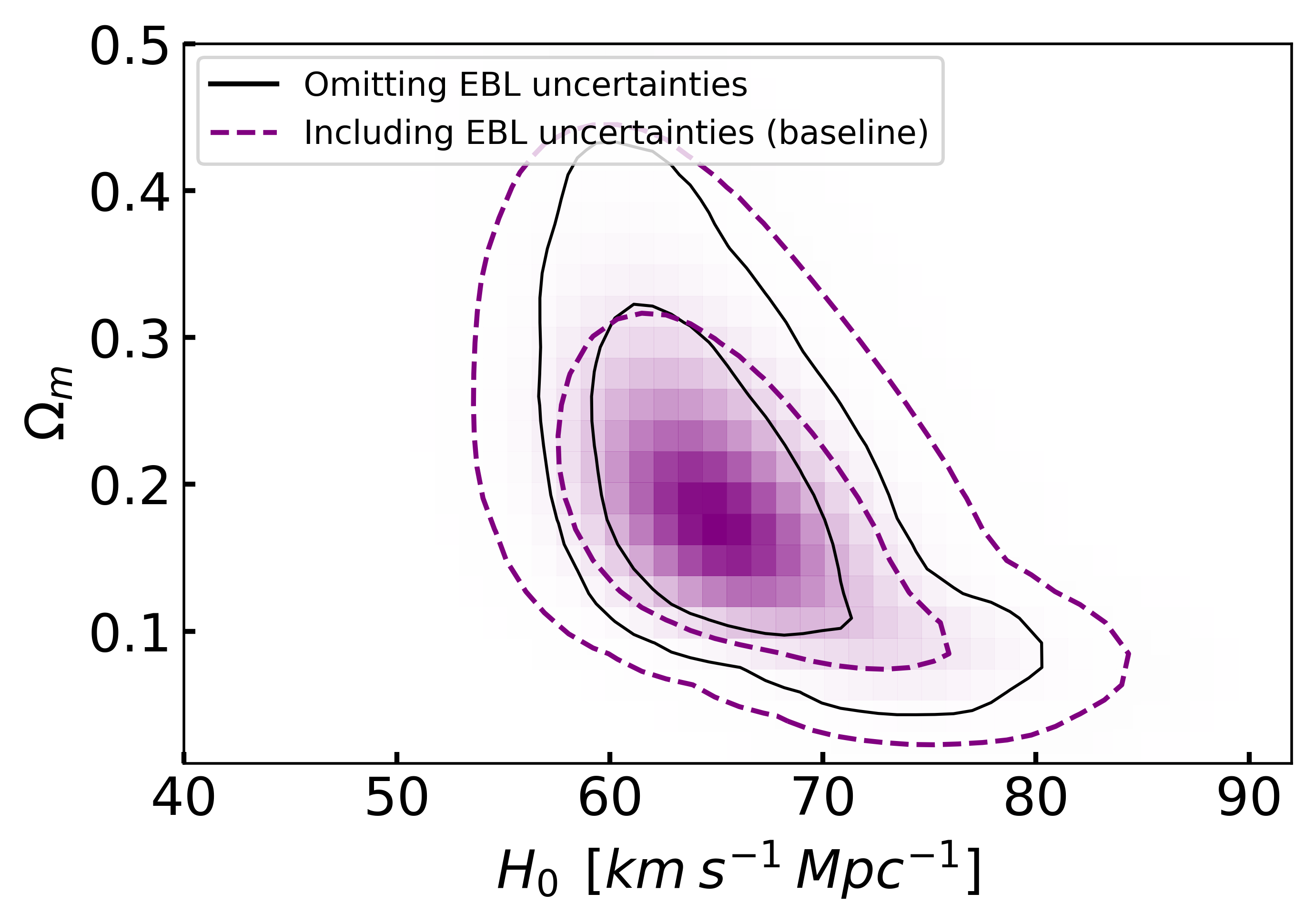}
    \caption{
    Impact of the uncertainties in the EBL model on cosmological constraints. The contours show constraints on the Hubble constant $H_{0}$ and the matter density parameter $\Omega_{\rm m}$ obtained in analyses including (the adopted approach in this work) or neglecting the EBL uncertainties (dashed purple line and solid black line, respectively). By neglecting the EBL uncertainties, the final errors in $H_{0}$ and $\Omega_{\rm m}$ decrease by $\sim 30$ and $\sim 10$ per cent, respectively.
    } 
    \label{fig:cosmo-EBL-errors}
\end{figure}

Noticeably lower estimates of the matter density parameter were found in previous analyses of the same $\gamma$-ray attenuation data, but using earlier versions of EBL models, $\Omega_{m}=0.14^{+0.06}_{-0.07}$ \citep{dominguez19}. Our study demonstrates that those underestimations resulted most likely from insufficient accuracy of the employed EBL models. Using the updated EBL model from \citetalias{saldana-lopez21} restores fair consistency with the matter content in the standard cosmological model; however, the previous trend with low values of $\Omega_{\rm m}$ seems to persist to some extent. It is natural to relate low best fit $\Omega_{\rm m}$ to the apparent discrepancy between the EBL model and $\gamma$-ray observations discussed in Section~\ref{sec:4.1}. Cosmological models with low matter density are characterized by larger distances at high redshifts. This primarily affects the line-of-sight integral given by Equation~\ref{attenu} and thus increases the optical depths at high redshifts to the level required by the $\gamma$-ray data, which would otherwise necessitate scaling up the EBL model for $\Omega_{\rm m}\approx0.3$ (see Figure~\ref{fig:ebl_evol}). The impact of high-redshift $\gamma$-ray data on cosmological constraints is clearly seen when we repeat the analysis in two separate redshift ranges. Figure~\ref{fig:cosmo-gamma} demonstrates explicitly that solutions with lower matter density are primarily driven by observations at higher redshifts, while fitting low-redshift yields cosmological constraints that are fully consistent with the Planck model. Note that (1) the $\gamma$-ray data uncertainties at the higher redshifts are large and (2) larger attenuation implies lower $\Omega_{m}$ \citep[see Figure 3 in][]{dominguez19}. The difference between the best fit $\Omega_{\rm m}$ measured from high- and low-redshift bins appears to be the most visible when splitting the data at $z=0.68$. Analysis of low-redshift data subset yields nearly the same estimate of the Hubble constant with comparable errors as for the entire data set, see Table~\ref{tab:cosmo-params}. This is also within 2$\sigma$ of the latest measurement of $H_{0}$ from the SH0ES program using observations of type Ia supernovae and Cepheids in relatively nearby galaxies, giving $H_{0}=73.3\pm 1.04\,\rm{km}\,\rm{s}^{-1}\,\rm{Mpc}^{-1}$ \citep{Riess2022}.

Cosmological measurements based on the $\gamma$-ray attenuation data can be further improved by combining them with independent cosmological probes giving complimentary constraints on $\Omega_{\rm m}$. Here, we use observations of the baryon acoustic oscillations (BAO) with the sound horizon normalisation set by the physical baryon density $\Omega_{\rm b}h^{2}$ obtained from the Big Bang nucleosynthesis theory. These are constrained by the local measurements of the primordial light element abundances, in particular $100\Omega_{\rm b}h^{2}=2.208\pm0.052$ derived from the recent determination of the primordial deuterium abundance in the most metal-poor damped Ly-$\alpha$ system \citep{Cooke2016}. We intentionally do not consider 
the Planck observations of the CMB or type Ia supernovae in order to keep our Hubble constant determination fully independent of those. For the BAO data we include anisotropic BAO measurements from the Sloan Digital Sky Survey's (SDSS) Baryon Oscillation Spectroscopic Survey \citep[consensus constraints based on post-reconstruction method][]{Ala2016}, distance measurements from the 6dF survey \citep{Beu2011} and from the Main Galaxy Sample of the SDSS \citep{Ros2015}. Cosmological constraints are obtained using the \textit{CAMB} code for cosmological calculations \citep{Lewis2000} and \textit{CosmoMC} \citep{Lewis2002} implemented in the \textit{cobaya} package \citep{Torrado2021} for computing Markov chains. We assume the CMB temperature measured by the COBE/FIRS, i.e. $T_{\rm CMB}=2.7255$~K \citep{Fixsen2009}. As shown in Figure~\ref{fig:cosmo-gamma-bao}, BAO observations 
and $\gamma$-ray attenuation data are complimentary in terms cosmological information. The apparent degeneracy between $H_{0}$ and $\Omega_{\rm m}$ from BAO data reflects a range of the sound horizon scale allowed by the adopted BBN prior. Combined analysis of both data sets yields stronger constraints on the Hubble constant than for the $\gamma$-ray data alone: $H_{0}=66.5^{+2.2}_{-2.1}\,\rm{km}\,\rm{s}^{-1}\,\rm{Mpc}^{-1}$ for the complete $\gamma$-ray data set and $H_{0}=69.8^{+3.0}_{-2.6}\,\rm{km}\,\rm{s}^{-1}\,\rm{Mpc}^{-1}$ for the low-redshift ($z<0.68$) subset, in both cases in agreement with the Planck value but in serious disagreement with local measurements assuming that systematic uncertainties in the EBL model do not exceed the quoted statistical errors. A slightly larger difference between these two estimates of the Hubble constant than in the case of including $\gamma$-ray data alone reflects relative shifts of the posterior probability contours from the BAO and $\gamma$-ray data on the $\Omega_{\rm m}-H_{0}$ plane (see Fig.~\ref{fig:cosmo-gamma-bao}).

Table~\ref{tab:cosmo-params} summarizes the results of our cosmological analysis. The constraints on the Hubble constant are consistent across the four data sets including low-redshift ($z<0.68$) $\gamma$-ray observations and combinations with the BAO measurements. In particular, virtually no difference (merely $\sim 0.5\sigma$ shift) between the full and low-redshift data in this respect demonstrates that the discrepancy between the EBL model in the near-infrared range and the $\gamma$-ray observations described in 
Section~\ref{sec:4.1} has a negligible impact on the Hubble constant estimation. Although the errors in our $H_{0}$ determination are about $3-4$ times larger than the total error quantifying the Hubble constant tension, our results clearly favor a Planckian value of $H_{0}$: the best fit 
$H_{0}$ values derived respectively from the full and low-redshift $\gamma$-ray data are merely $(0.5-1.0)\sigma$ lower than the Planck value \citep{planck19}, but $(1.6-2)\sigma$ lower than the local $H_{0}$ from SH0ES \citep{Riess2022}. Similar preference for the Planck value of $H_{0}$ was reported in previous studies based on 
different EBL models \citep{dominguez19}. We note that another group, the Carnegie Supernova Project, finds a different $H_{0}$ value with Type Ia supernova \citep{freedman19}.

Uncertainties on the obtained cosmological constraints is primarily driven by the errors of the $\gamma$-ray attenuation measurements. This is demonstrated by Figure~\ref{fig:cosmo-EBL-errors} that compares constraints on $H_{0}$ and $\Omega_{\rm m}$ from complete $\gamma$-ray observations conditioned to the best fit EBL model or employing marginalisation over the allowed EBL models. It is apparent that including the EBL errors in the analysis increases the contours only marginally. The corresponding errors in $H_{0}$ and $\Omega_{\rm m}$ increase respectively by $\sim 30$ and $\sim 10$ per cent relative to the results of the analysis neglecting the EBL errors: $H_{0}=64.5^{+4.6}_{-3.3}$~km~s$^{-1}$~Mpc$^{-1}$ and $\Omega_{\rm m}=0.19\pm0.07$.


\section{Summary and conclusions}\label{sec:conclusions}
This work builds upon a recently published EBL model by \citetalias{saldana-lopez21} that focuses on improving the estimates of the spectral intensity evolution at higher redshifts and the IR region. We derive optical depths that can be used for correcting EBL-attenuated spectra observed with $\gamma$-ray telescopes. For instance, given the improvements in the derivation of the mid-to-far IR peak of the EBL and its evolution over redshift from \citetalias{saldana-lopez21}, our optical depths are suited for spectra at high redshifts and also up to the highest energies. This makes them optimal for observations with {\it Fermi}-LAT, the current generation of IACTs, and specially the future Cherenkov Telescope Array (CTA).

We find that the optical depths derived from $\gamma$-ray attenuation \citep{ebl18,desai19} agree within 2$\sigma$ with those derived from \citetalias{saldana-lopez21} using galaxy data. A comparison of these optical depths from data and model allows us to measure $H_{0}$ and $\Omega_{m}$, finding that $H_{0}$ is $1\sigma$ compatible with the value obtained from cosmological probes such as the CMB and BAO, whereas $\Omega_{m}$ is 2$\sigma$ compatible. This marginally low value of $\Omega_{m}$ is produced by the attenuation obtained from higher redshift blazars. A careful study on the propagation of the EBL uncertainties to the cosmological parameters is also developed. Finally, we note that a larger EBL intensity in the model will make $H_{0}$ and $\Omega_{m}$ derived from $\gamma$-ray attenuation larger. Given the recent results by \citet{lauer22} of an optical intensity that is approximately 50\% larger than the one derived galaxy counts \citep[e.g.,~][]{driver16}, the problem of how much light there is in the Universe remains puzzling.



\section*{Acknowledgements}

We thank Daniel Nieto for providing computational resources, Jonathan Biteau for helpful comments, and the anonymous referee for the review. A.D. is thankful for the support of the Ram{\'o}n y Cajal program from the Spanish MINECO, Proyecto PID2021-126536OA-I00 funded by MCIN / AEI / 10.13039/501100011033, and Proyecto PR44/21‐29915 funded by the Santander Bank and Universidad Complutense de Madrid. RW was supported by a grant from VILLUM FONDEN (project number 16599). ASL acknowledge support from Swiss National Science Foundation. J.F. was supported by NASA through contract S-15633Y and through the Fermi GI program. PGP-G acknowledges support from  Spanish  Ministerio  de Ciencia e Innovaci\'on MCIN/AEI/10.13039/501100011033 through grant PGC2018-093499-B-I00.


\section*{Data Availability}

The results of this study are publicly available at \url{https://www.ucm.es/blazars/ebl} and by request to the authors.



\bibliographystyle{mnras}
\bibliography{mybiblio} 








\bsp	
\label{lastpage}
\end{document}